**High-Speed and Continuous-Wave Programmable Luminescent Tags Based on Exclusive Room Temperature Phosphorescence (RTP)**

*Max Gmelch, Tim Achenbach, Ausra Tomkeviciene, and Sebastian Reineke\**

Dr. Max Gmelch, Tim Achenbach, Dr. Ausra Tomkeviciene, Prof. Dr. Sebastian Reineke
Dresden Integrated Center for Applied Physics and Photonic Materials (IAPP) and Institute for Applied Physics, Technische Universität Dresden, 01187 Dresden, Germany
E-mail: sebastian.reineke@tu-dresden.de

Dr. Ausra Tomkeviciene
Department of Polymer Chemistry and Technology, Kaunas University of Technology, K. Barsausko g. 59, 51423 Kaunas, Lithuania

Keywords: room temperature phosphorescence, RTP, programmable tags, photo-induced optical memory, smart materials

**Abstract**

Most materials recently developed for room temperature phosphorescence (RTP) lack of practical relevance due to their inconvenient crystalline morphology.[1–5] Using amorphous material systems instead, programmable luminescent tags (PLTs) based on organic biluminescent emitter molecules with easy processing and smooth sample shapes were presented recently.[6,7] Here, the effective quenching of the emitter's RTP by molecular oxygen ($O_2$) and the consumption of the excited singlet $O_2$ through a chemical reaction represent the central features. With customized activation schemes, high resolution content can be written and later erased multiple times into such films, providing a versatile yet simple photonic platform for information storage. However, two important limitations remain: (i) The immutable fluorescence of the emitters outshines the phosphorescent patterns by roughly one order of magnitude, allowing read-out of the PLTs only after the excitation source is turned off. (ii) The programming of these systems is a rather slow process,[8–13] where lowest reported activation times are still > 8 s.[6] Here, a material-focused approach to PLTs with fast activation times of $120 \pm 20$ ms and high-contrast under continuous-wave (cw) illumination is demonstrated, leading to accelerated programming on industry relevant time scales and a simplified readout process both by eye and low cost cameras.

**Main Text**

RTP from organic molecules receives plenty of attention recently.[1,14–17] While crystallization-induced phosphorescence is the main focus, RTP from amorphous emitters shows higher flexibility in sample processing and applications.[18,19] The proposed applications reach from afterglow security features[10–12,20,21] to thin film oxygen sensors.[22–26] The latter make use of the fact that excited triplet states in organic molecules are efficiently quenched by molecular oxygen via triplet-triplet-interactions (TTI), leading to nonradiative relaxation of the emitter and generation of excited singlet state oxygen.[27] In this electronic configuration, oxygen is highly reactive.[28] In a polymer-based film, ongoing generation of singlet oxygen can therefore lead to a total vanishing of molecular oxygen through so called photo consumption, i.e., the oxidation of the polymer.[29]

Unwanted in organic oxygen sensors,[30] this effect opens the possibility to activate phosphorescence in transparent emitter-doped polymer films locally and therefore to print any pattern into the layers by mask illumination (**Figure 1**a).[6] Subsequently, this imprint can be read out by monitoring the phosphorescence emerged at the irradiated locations with reduced excitation intensity (Figure 1b). By heating via infrared light or a hotplate, the patterns can be fully erased again, since the oxygen permeability of an oxygen barrier material, coated on top of the emission layer, is increased,[31] and the emission layer is repopulated with molecular oxygen (Figure 1c). Consequently, PLTs with multiple reading, writing, and erasing cycles are realized. Though, in cw illumination, the relatively weak phosphorescence is overlaid by the dominating fluorescent emission of the emitter N,N′-di(1-naphthyl)-N,N′-diphenyl-(1,1′-biphenyl)-4,4′-diamine (NPB, **Figure S1**). Hence, in order to read the imprinted structures, accurate timing is required to allow the detection of the patterns as a short afterglow emission right after turning off the excitation source (c.f. Figure 3d). For implementation in real life applications, this time-critical procedure is not suitable.

With regard to the realization of high-contrast PLTs with readability under cw illumination (cw-PLTs), one can therefore formulate several desirable properties of the material system. On the one hand, the host material should be (I) amorphous and transparent, (II) capable of storing oxygen, (III) sensitive to oxidation, and, (IV) once oxygen is vanished, no fresh oxygen may diffuse into the emitting layer. On the other hand, the desired emitter (V) is showing RTP with high quantum yield ($\phi_P$) in the absence of molecular oxygen, (VI) is sensitive to efficient quenching of RTP in the presence of molecular oxygen, and, as already mentioned, (VII) its fluorescence quantum yield ($\phi_F$) should be very low compared to $\phi_P$.

While the host material requirements I to IV are already addressed adequately in our previous work using poly(methyl methacrylate) (PMMA) as host[6,29] and a modified ethylene-vinyl alcohol copolymer as oxygen barrier,[6] realizing all of the listed requirements I to VII simultaneously is not trivial.

Emitters containing heavy atoms, like iridium or platinum, do on the one hand show high $\phi_P$. Further, their fluorescence is vanished completely through high intersystem crossing caused by the increased spin-orbit coupling.[32] On the other hand, the influence of oxygen is weak to negligible (cf. Figure 4b), since the quenching rate caused by TTI is lower or in the same order as the radiative rate from the triplet. Embedded in PMMA at low concentration and in the absence of oxygen, we measured values for $\phi_P$ of 41% and 19% for Ir(MDQ)$_2$(acac) and PtOEP, respectively. At ambient conditions, values of 41% and 10% were reached (c.f. Tab 1). Hence, PtOEP emission is enhanced by a factor of only 1.6 in oxygen-free atmosphere, and Ir(MDQ)$_2$(acac) remains fully unchanged. The same accounts for its phosphorescence lifetime $\tau_{IrMDQ}$ of $1.5 - 1.6$ µs, while for PtOEP it changes from $\tau_{PtOEP,O2} = 31$ µs to $\tau_{PtOEP,N2} = 53$ µs going from ambient to inert atmosphere. Consequently, both materials are not well suited for the realization of cw-PLTs. Nevertheless, these data give access to the average oxygen quenching rate $k_{O2}$ in the system via the following equation (see Supp. Inf. for derivation):

$$k_{O2} = \frac{1}{\tau_{PtOEP,O2}} - \frac{1}{\tau_{PtOEP,N2}} = 1.3 \times 10^4 \frac{1}{s} \tag{1}$$

Considering this number as a rough generalized value for oxygen quenching of any emitter embedded in PMMA, a lower phosphorescent lifetime limit for sufficient quenching of phosphorescence ($\tau_{O2} \overset{!}{<} 0.01 \times \tau_{N2}$) can be denoted (see SI for derivation):

$$\tau_{N2} > 10^{-2} s = 10 \text{ ms} \tag{2}$$

Together with $\phi_P$, this can be used as benchmark value for the requirements V and VI. For VII, the ratio of $\phi_P$ to $\phi_F$ is used, with higher values representing better contrast of the tags:

$$P2F = \frac{\Phi_P}{\Phi_F} \tag{3}$$

Checking on three recent reviews[32–34] containing more than 290 RTP systems, not one of them fulfils all requirements sufficiently, albeit two of them are worth a closer look. First, Hirata et al. realized a system showing efficient phosphorescence with high $\phi_P = 12.1\%$.[35] The steroidal host material prevents oxygen from quenching the phosphorescence, which is beneficial for applications requiring invariable emission. However, at the same time it hinders spatially resolved oxygen quenching in the emitting layer, as would be required for realizing luminescent tags. Further, its P2F = 0.8 is insufficient for high luminescent contrast. Chen et al. realized P2F values up to > 39, using donor-acceptor (D-A) charge transfer (CT) states localized on the same molecule.[36] In detail, the authors enhanced the triplet formation processes by reducing the singlet-triplet splitting energy $\Delta E_{ST}$ via forming CT states. The high P2F value, though, is not caused by intense phosphorescence, but originates from the almost complete vanishing of the fluorescence, while the RTP efficiency did not exceed $\phi_P = 3.9\%$.

Still, the utilization of CT states as intermediate level for populating local excited triplet states ($^3$LE), is a promising approach for increasing $\phi_P$, since it is known that excited triplet state generation might be a favorable energetic pathway after CT state forming.[37] Consequently, a D-A molecule comprising a donor or acceptor unit with intrinsically efficient $^3$LE emission may lead to an increase of the P2F value.

Here we used thianthrene (TA) as donor units, which shows notable RTP both in crystalline state[38] and amorphous films,[39] and benzophenone (BP) as electron acceptor. The resulting emitter 4,4'-dithianthrene-1-yl-benzophenone (BP-2TA), which we recently synthesized as part of a series of D-A materials for oxygen sensing,[40] comprises two TA units linked to BP via C-1 position (**Figure 2**a). This linkage minimizes the conjugation between the donor and acceptor units. Embedded in amorphous PMMA at low concentration (0.5 wt% to 20 wt%) and illuminated with 340 nm, the system shows weak emission in the presence of oxygen (Figure 2b), while purging with nitrogen enables phosphorescence with high $\phi_P$ up to 21% (Figure 2c). Despite a small decrease in peak energy, the spectral shape of the BP-2TA phosphorescence resembles the pure TA $^3$LE emission (Figure 2d). This hints to the donor's local excited triplet state being the origin of the BP-2TA triplet emission. Comparable behavior was observed for different TA derivatives similar to BP-2TA.[41]

The redshift of the fluorescence of BP-2TA compared to the prompt TA and BP emission (Figure 2e), as well as the comparison of theoretical calculations[40–42] (Figure 2f) verify the presence of a prompt emitting CT state.

A comparison of $\phi_P$ of pure TA (6%) and BP (~3%) (Figure 2c) to BP-2TA reveals a drastic phosphorescence yield increase when combining donor and acceptor to one molecule. At the same time, the fluorescence CT state yield of BP-2TA $\phi_F$ = 1% to 1.5%, depending on the concentration, is reduced compared to pure TA fluorescence yield (3%). This implies a successful realization of efficient $^3$LE population via an intermediate CT state.

Interestingly, excitation scans of BP-2TA show different maxima for CT fluorescence ($\lambda_{max,exc}$ = 340 nm) compared to $^3$LE phosphorescence ($\lambda_{max,exc}$ = 320 nm), monitored at the same emission wavelength ($\lambda_{em}$ = 470 nm, **Figure S2**). This leads to an excitation-wavelength dependent P2F value, which (at 5 wt%) drops from 18 to 8 when going from $\lambda_{exc}$ = 340 nm to $\lambda_{exc}$ = 365 nm (**Figure S3**). However, this still is a suitable value for high contrast patterns (cf. **Figure 3**c and Figure 4c).

In total, with high $\phi_P$ and P2F, low $\phi_F$, and an intensity-averaged phosphorescence lifetime of $\tau$ = 30 ms, BP-2TA fulfils all criteria for realizing high contrast cw-PLTs. To do so, tags were prepared as described in our previous publication,[6] consisting of a PMMA:BP-2TA emitting layer, coated with an oxygen barrier material on top (Figure 3c top right). Since prepared in ambient atmosphere, the bottom layer, despite being protected from penetration of exterior one, still contains molecular oxygen. Therefore, in a fresh sample, phosphorescence is fully quenched by TTI, leading to very weak luminescent emission (Figure 3a, first seconds). In contrast to that, ongoing illumination at some point leads to the emergence of intense triplet emission, since oxygen in the emitting layer vanishes through photo consumption (Figure 3a). In this manner, any phosphorescent pattern can be printed into the layer via mask illumination. Since the printing time is inversely proportional to the UV intensity (Figure 3b), it can be reduced to $120 \pm 20$ milliseconds by using a high-intensity UV light source (**Figure S4**). Reading the imprint is now, in contrast to prior publications,[6,7] easily possible in low-intensity cw-illumination without precise timing, because there is no dominating fluorescence overlaying the triplet emission (Figure 3d).

By heating the tag on a hotplate at 90°C for a few minutes, the phosphorescence is erased due to oxygen refilling. Subsequently, new patterns can be printed multiple times (Figure 3c).

For consolidation of the presented design concept, in **Figure 4** and Table 1, different RTP emitters, spanning from NPB, Ir(MDQ)$_2$(acac) and PtOEP to BP-2TA, with different $\phi_P$, $\phi_F$, phosphorescence lifetimes and P2F values, are compared regarding their suitability for cw-PLTs.

In conclusion, we present a general approach to realize high-speed writeable and high-contrast cw-readable luminescent tags, which was successfully applied using a highly-efficient amorphous RTP emitter, showing cw-suitable PLTs for the first time. The cw-reading capabilities open up the possibilities to an easy data readout using the bare eye or a simple smartphone camera. The fast printing mechanism using high-intensity UV light guaranties a fluent processing. Further development of amorphous RTP materials fulfilling the PLT criteria is necessary.

**Experimental Section**

*Materials:* PMMA (average molecular weight 550,000) was purchased from Alfa Aesar. BP was purchased from Sigma Aldrich, TA from TCI Deutschland GmbH, NPB and Ir(MDQ)$_2$(acac) from Lumtec Technology Corp. and PtOEP from Frontier Scientific. All these materials were used without further purification. The synthesis of BP-2TA has been described in detail before.[40] The oxygen-barrier material was purchased from Kuraray Europe GmbH and contains modified ethylene-vinyl alcohol copolymers.

*Film fabrication:* PMMA and the respective emitter both were dissolved in anisole to get a solution containing 95 wt% PMMA and 5 wt% of the emitter. The oxygen barrier material was dissolved in distilled water at 150°C. The layers were either spin coated at a speed of 2000 rpm using 150 μl of the solution, or drop casted using 500 μl on 1-inch quartz glass substrates. Emitting and oxygen-barrier layer were coated on top of each other with enough storing time to fully dry in-between.

*Lifetime measurements:* Lifetimes in the ms-regime were determined using a 340 nm M340L4 LED or a 365 nm LED M365L2 (Thorlabs), a TGP3122 pulse generator (AIM-TTI Instruments), and a silicon photodetector PDA100A (Thorlabs). For the ns- and μs-regime, a Time Correlated Single Photon Counting (TCSPC) Setup was used, containing of a 375 nm laser LDHDC375, a PMA Hybrid Detector PMA Hybrid 40, a TimeHarp platine (all PicoQuant), and a Monochromator SpectraPro HRS-300 (Princeton Instruments). ns-Lifetimes were evaluated using reconvolution algorithms of FluoFit (PicoQuant).

*Spectral measurements:* Prompt emission spectra were recorded using the LEDs mentioned above and a Spectrometer CAS 140CT-151 (Instrument Systems). For delayed phosphorescence spectra, the LED and the spectrometer were triggered, resulting in spectra acquired shortly after the LED turned off.

*Activation curves and activation time:* To get the phosphorescence activation curves, spectra were recorded subsequently. The emission intensity then was integrated over the whole visible range and plotted against the illumination time. The activation time of the low-intensity measurements was derived from the activation curves, for the phosphorescence reaching half of the maximum measured phosphorescence. The derivation of the activation time of the high-intensity measurement is described in details in Figure S4.

*PLQY measurements:* To determine the photoluminescent quantum yields of the different materials, an improved procedure[43] following de Mello's method[44] was used. For fluorescence PLQY, the samples were measured in ambient atmosphere. For phosphorescence PLQY, if not marked otherwise, nitrogen atmosphere was used. Subsequently, the phosphorescence PLQY was determined by subtracting the fluorescence PLQY value from the measurement result. For very low fluorescence signals, the fluorescence PLQY was determined by a spectral analysis of a spectrum containing both fluorescence and phosphorescence.

*Excitation scan measurements:* Excitation scans were performed with a SPEX FluoroMax (Horiba).

*Mask illumination:* Masks were printed with an office printer onto common overhead transparencies and then put in the UV beam path. Subsequently, the LED illuminated the sample through the mask, resulting in special resolved activation of phosphorescence. Before recording photos of the cw-PLTs, the masks were removed.

*High-intensity UV radiation:* As high-intensity UV light source, a 365 nm LED M365LP1 (Thorlabs) and a focusing lens ACL25416U (Thorlabs) were used, reaching an intensity of 2.3 W cm$^{-2}$.

*Photographs:* A conventional digital camera EOS60D from Canon was used to take the photographs. Some images are slightly contrast corrected to ensure sufficient image quality when printed.

*Sample heating:* To erase the phosphorescence, a common hotplate was used.

**Supporting Information**

Supporting Information is available from the Wiley Online Library or from the author.

**Acknowledgements**

This project has received funding from the European Research Council (ERC) under the European. Union's Horizon 2020 research and innovation program (grant agreement No 679213 "BILUM"). The Authors thank Toni Bärschneider, Felix Fries, Anton Kirch and Paulius Imbrasas for fruitful discussions.

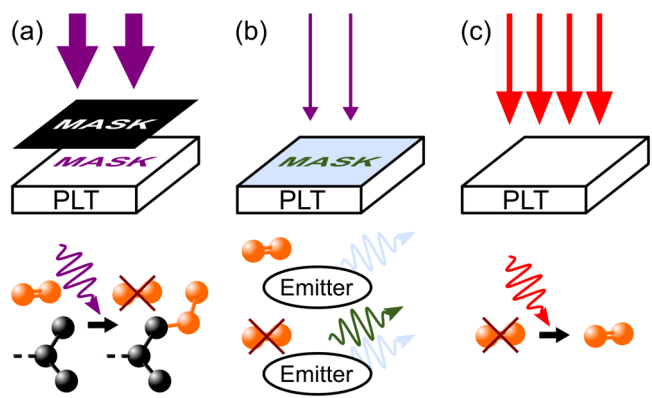

**Figure 1.** (a) Writing: through mask illumination with high-intensity UV radiation (broad violet arrows), molecular oxygen (orange) in the emitting layer undergoes photo consumption and therefore vanishes in the irradiated areas by oxidizing the host polymer (black, only sketched partly). (b) Reading: after removing the mask, only the imprinted areas show phosphorescent emission (green) due to the absence of molecular oxygen. Additionally, the whole PLT shows unwanted fluorescence (blue), independent of oxygen concentration. To avoid further activation of phosphorescence, the UV intensity is reduced (small violet arrows). (c) Erasing: by heating the PLT using a hotplate or infrared light (red arrows), the emitting layer is refilled with oxygen, leading to a vanishing of the phosphorescence. This cycle is repeatable multiple times.

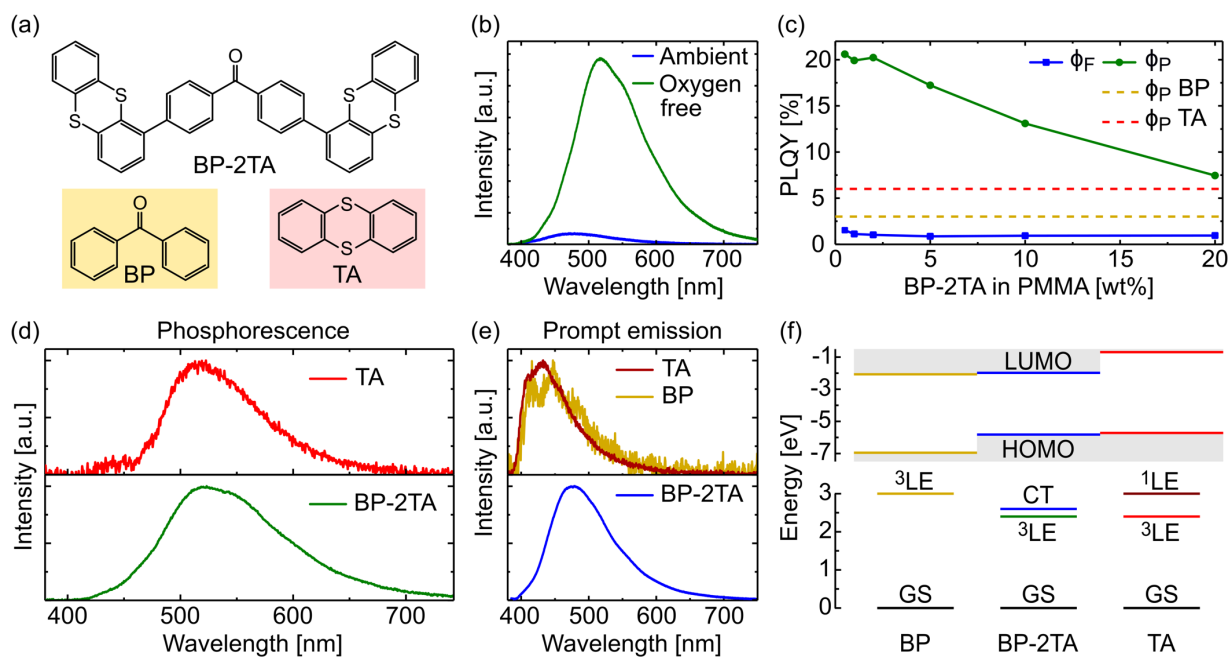

**Figure 2.** (a) Molecular structure of BP-2TA (not shaded), BP (yellow shaded) and TA (red shaded). (b) Emission of a drop-casted PMMA:BP-2TA (5 wt%) sample in ambient conditions (blue) and nitrogen atmosphere (green) at room temperature. Excitation wavelength was 340 nm. (c) Phosphorescence (green) and fluorescence (blue) quantum yield of samples with different concentrations of BP-2TA in PMMA. $\phi_P$ of PMMA:BP (5 wt%) (yellow) and PMMA:TA (2 wt%) (red) are shown as comparison, measured in nitrogen atmosphere. Excitation wavelength was 340 nm. (d) Phosphorescence spectra of TA (red) and BP-2TA (green), measured shortly after excitation turnoff. (e) Prompt emission of TA (dark red), BP (dark yellow) and BP-2TA (blue). Note that while TA emission is fluorescence, BP emission consists mainly of fast phosphorescence[45]. (f) Top: highest occupied molecular orbit (HOMO) and lowest unoccupied molecular orbit (LUMO) energies of BP (dark yellow), BP-2TA (blue) and TA (red), with values taken from different publications[40–42]. Bottom: Excited state diagram, calculated from emission peak energies, showing the BP triplet $^3$LE (dark yellow), TA singlet $^1$LE (dark red) and triplet $^3$LE (red), as well as BP2-TA CT (blue) and $^3$LE (green) state, referring to the respective ground state GS (black).

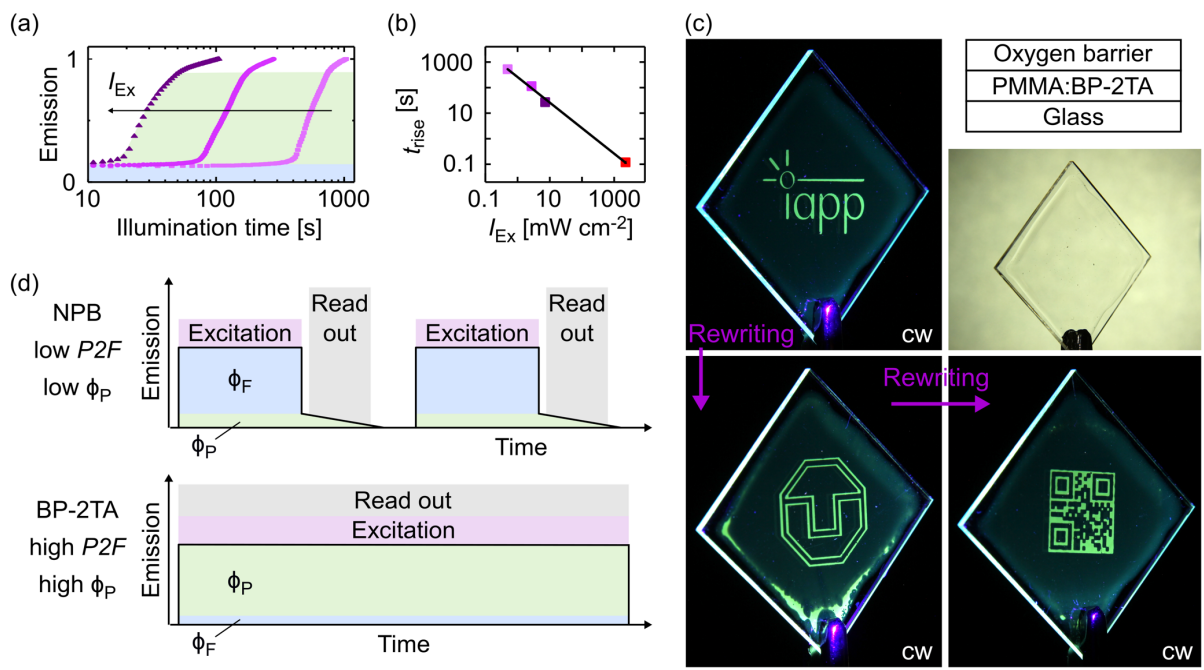

**Figure 3.** (a) Activation curve of PMMA:BP-2TA (5 wt%), showing an emission increase due to the activation of phosphorescence after a certain illumination time for different 365 nm LED intensities, ranging from 0.5 mW cm⁻² to 7.2 mW cm⁻². (b) Illumination time until phosphorescence arises, dependent on the excitation intensity. The purple dots emerge from the graph in b; for the derivation of the high intensity value (red dot) see Fig S4. The black line is a power-law fit using an exponent of -1. (c) A PLT containing PMMA:BP-2TA (5 wt%), which is transparent in visible light, was erased twice by heating to 90°C for 3 minutes and 6 minutes, respectively. Consequently, new patterns were printed by mask illumination using a 365 nm LED as published before[6]. For readout, a 365 nm LED with reduced intensity was used in cw mode. Top right: Sample structure of the PLT. (d) PLT readout scheme for tags from previous publications[6,7] (top) with high $\phi_F$ and low $\phi_P$, including a complex timing routine for excitation and readout, compared to the new method (bottom), where, due to high $\phi_P$ compared to $\phi_F$, read out is feasible without any timing procedures.

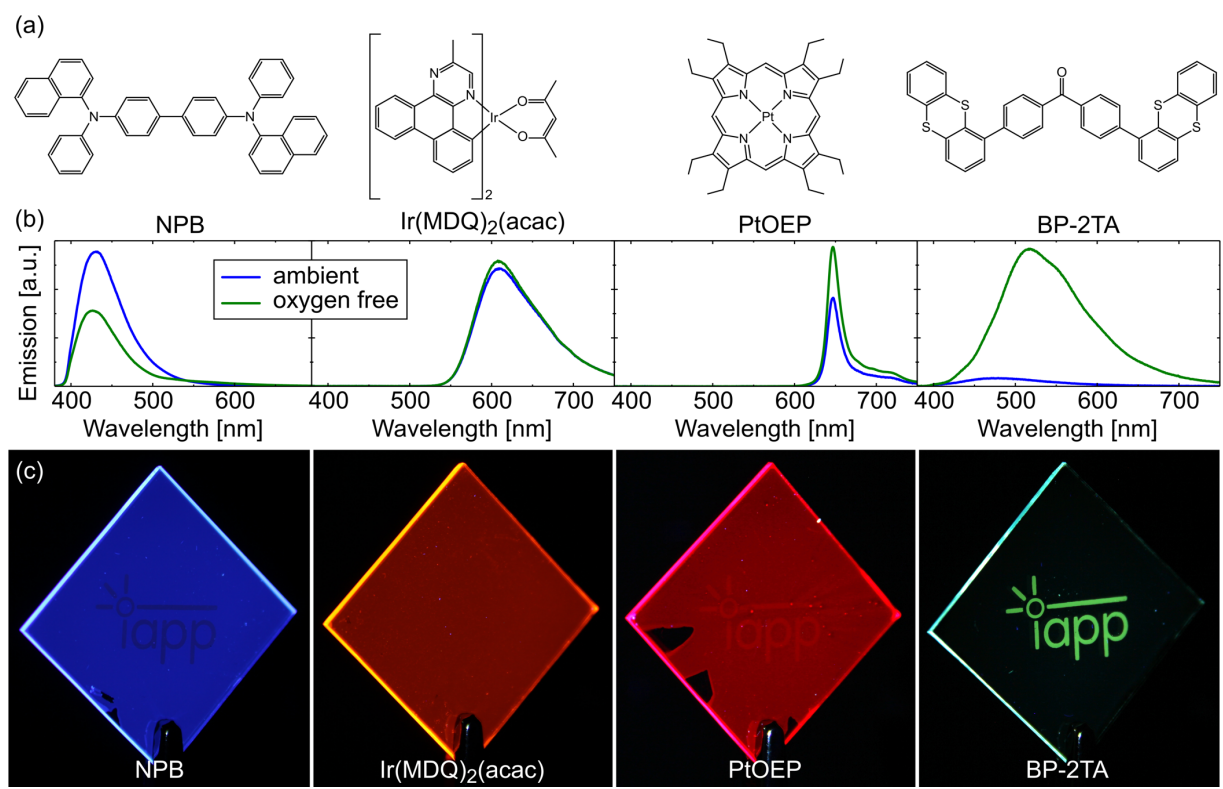

**Figure 4.** (a) Molecular structure of NPB, Ir(MDQ)₂(acac), PtOEP and BP-2TA. (b) Corresponding emission spectra in ambient and inert nitrogen atmosphere, recorded from samples containing the emitters diluted into PMMA at identical molecular concentrations. While conventional emitters show little to no increase in intensity upon the removal of oxygen, BP-2TA emission increases notably. The decrease of NPB emission results from excited state annihilation due to the high triplet density in this system. (c) cw-PLT realization using the respective emitters. While Ir(MDQ)₂(acac) shows no image and PtOEP a barely visible one, BP-2TA reveals a high contrast imprint in cw illumination. In NPB, at high intensity, a very weak inverse pattern is visible, again due to excited state annihilation in the presence of triplets. Excitation wavelength was 365 nm, except for BP-2TA, where 340 nm was chosen.

| emitter | $\tau_F$ (ns) | $\tau_P$ (ms) | $\phi_F$ (%) | $\phi_P$ (%) | P2F | V | VI | VII |
|---|---|---|---|---|---|---|---|---|
| NPB | 3.2 | 348 | 22[a] | ~3[b] | <<1 | - | X | - |
| Ir(MDQ)$_2$(acac) | - | 0.002 | - | 41 | very high[c] | - | - | X |
| PtOEP | - | 0.031/0.053[d] | - | 10/19[d] | very high[c] | - | - | X |
| BP-2TA | 3.3 | 30 | 1 | 18/21[e] | 18/21[e] | X | X | X |

**Table 1.** Different RTP emitter properties and their aptitude for cw-readout PLTs.

$\tau_F$: intensity-weighted average fluorescence lifetime

$\tau_P$: intensity-weighted average phosphorescence lifetime

$\phi_F$: fluorescence quantum yield

$\phi_P$: phosphorescence quantum yield

P2F: ratio of phosphorescence to fluorescence

V, VI, VII: respective criterion, referring to the text

X: fulfilling the respective criterion

a: in the absence of excited triplet states

b: for low excitation density

c: no exact number, since fluorescence is not measurable

d: in ambient/inert atmosphere

e: for high (5 wt%) / low (0.5 wt%) concentration and excitation at 340 nm

Using organic room temperature phosphorescence (RTP), programmable luminescent tags (PLTs) enable multiple writing, reading and erasing of information into or from transparent polymer films. Here, an overview of crucial parameters and material properties for the realization of PLTs is given and consecutively implemented, leading to high-contrast PLTs with fast writing and simplified reading procedures.

**Keyword**

room temperature phosphorescence

**Authors**

Max Gmelch, Tim Achenbach, Ausra Tomkeviciene, and Sebastian Reineke*

**Title**

High-Speed and Continuous-Wave Programmable Luminescent Tags Based on Exclusive Room Temperature Phosphorescence (RTP)

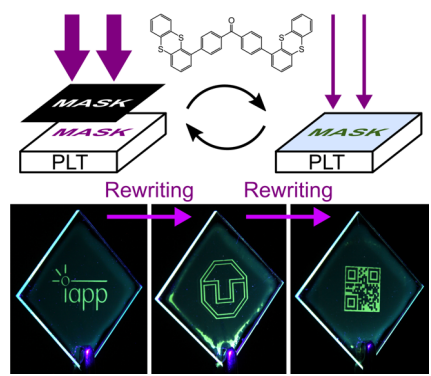

# Supporting Information

**Supplementary Information – High-speed and continuous-wave programmable luminescent tags**

*Max Gmelch, Tim Achenbach, Ausra Tomkeviciene, and Sebastian Reineke\**

**Derivation of the average oxygen quenching rate $k_{O2}$**

$$\tau_{O2} = \frac{1}{k_r + k_{nr} + k_{O2}} \tag{S1}$$

$$\tau_{N2} = \frac{1}{k_r + k_{nr}} \tag{S2}$$

$$\frac{1}{\tau_{O2}} - \frac{1}{\tau_{N2}} = k_r + k_{nr} + k_{O2} - k_r + k_{nr} = k_{O2} \tag{S3}$$

While $\tau_{O2}$ is the phosphorescence lifetime in the presence of oxygen, $\tau_{N2}$ is the phosphorescence lifetime in the absence of oxygen, $k_r$ is the radiative and $k_{nr}$ the intrinsic nonradiative rate.

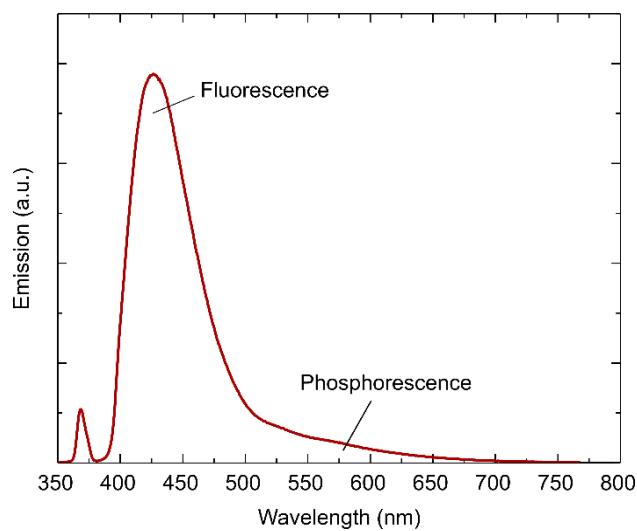

**Figure S1.** Emission of a PMMA:NPB (5 wt%) sample in nitrogen atmosphere. The weak phosphorescence at 550 nm is outcompeted by the intense fluorescence.

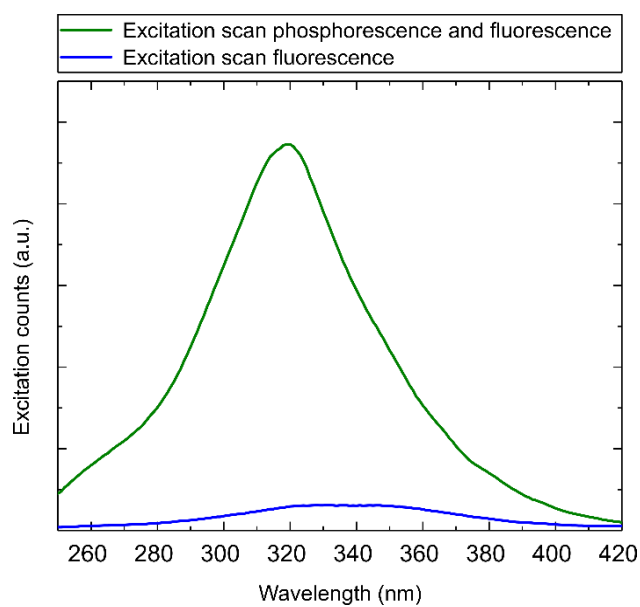

**Figure S2.** Excitation scans of PMMA:BP-2TA (5 wt%) The fluorescence excitation scan (blue) was recorded using a PLT sample before activating the phosphorescence, the combined phosphorescence and fluorescence scan (green) after activation of phosphorescence. For both measurements, the emission wavelength was set to 470 nm.

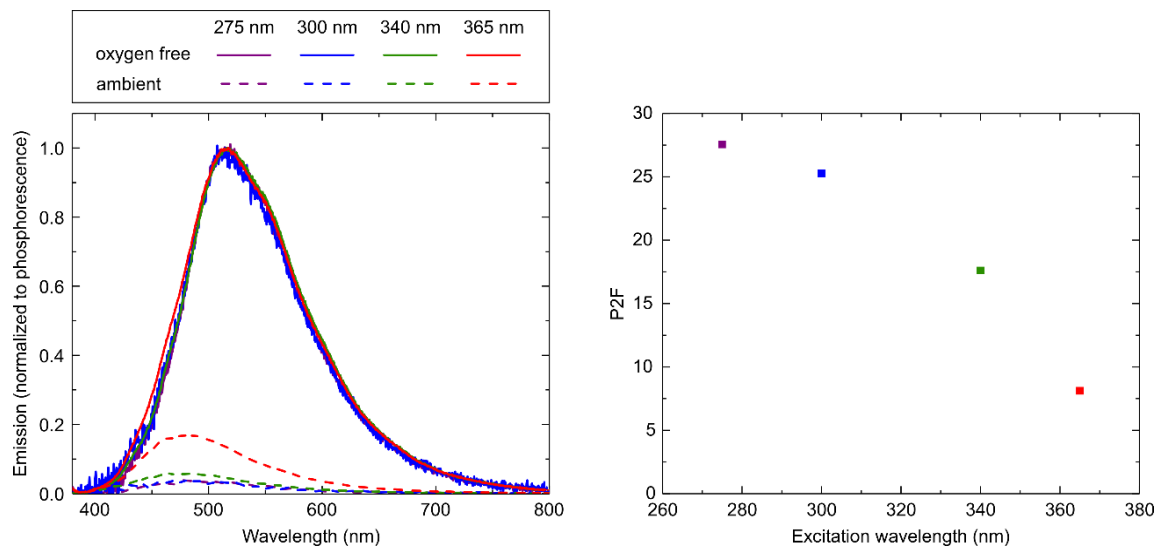

**Figure S3.** Excitation wavelength dependence of PMMA:BP-2TA (5 wt%). Spectra were taken in ambient and nitrogen atmosphere to measure fluorescence and phosphorescence. To calculate the P2F ratio, the spectra were integrated from 400 nm to 750 nm.

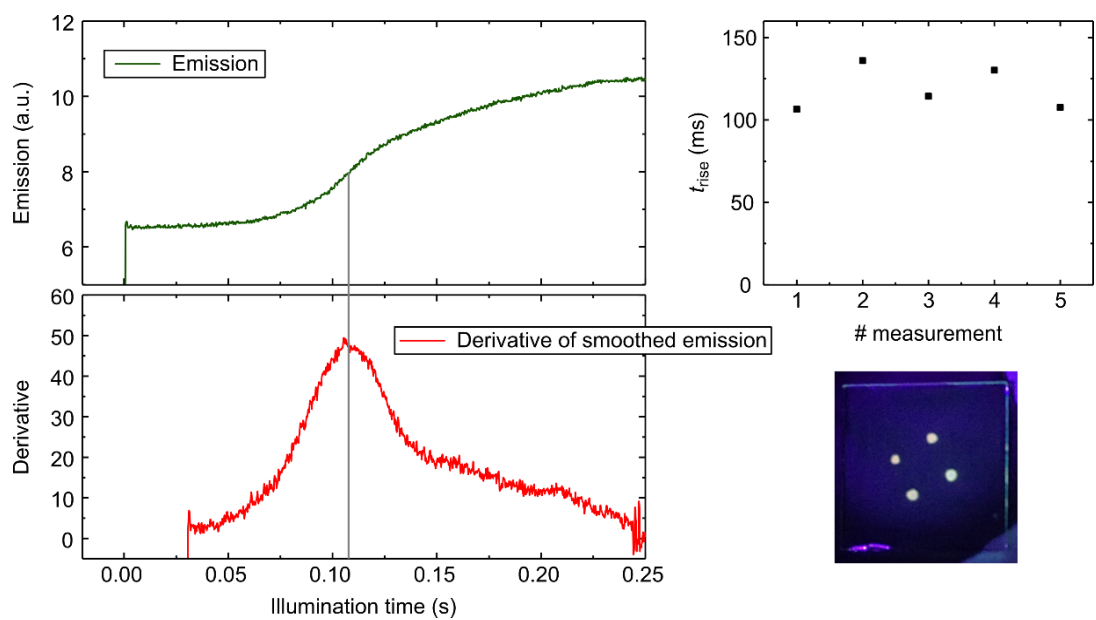

**Figure S4.** High-intensity UV activation of a cw-PLT containing PMMA:BP-2TA (5 wt%). Using a high-intensity UV source, a PLT was illuminated through a 1x1 mm² pinhole and the emission increase was monitored. Right behind the pinhole, the UV intensity was 2.3 W cm$^{-2}$. Due to scattered light, the sample got additionally partly illuminated in the area around the pinhole with lower intensity. Therefore, the measured rise of the emission did not stop sharply after activation of the area right behind the pinhole, leading to a smeared out plateau of the emission. Thus, for extracting the rise time, here the inflection point of the emission rise was taken as $t_{rise}$. To do so, the maximum of the first derivative was used. Five measurements were acquired, leading to a mean value of $t_{rise} = 120 \pm 20$ ms. The photograph shows four different measurements. The different spot sizes result from the activation of phosphorescence beyond the pinhole, leading to the smeared out plateau.

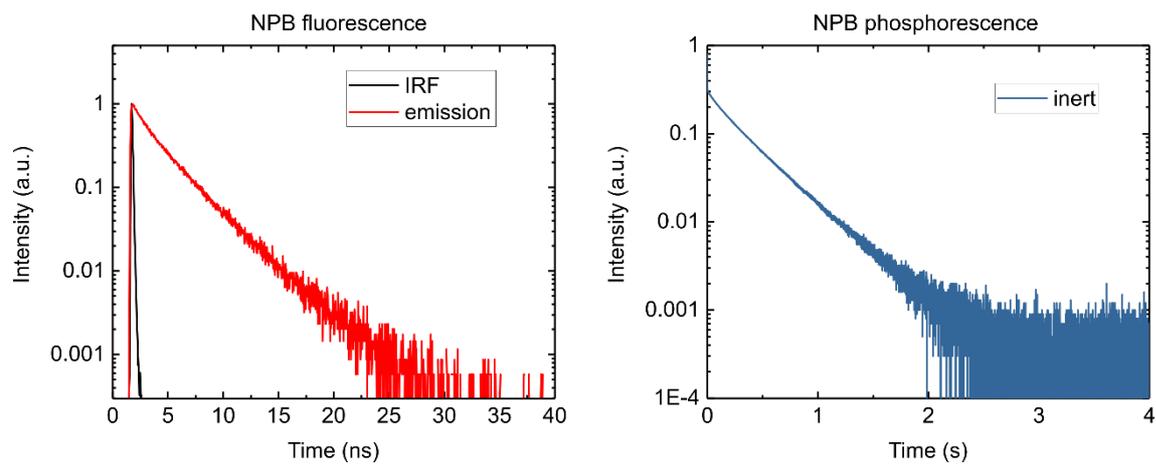

**Figure S5.** Fluorescence and phosphorescence decays of a PMMA:NPB (5 wt%) sample. The fluorescence decay was measured in ambient atmosphere; the phosphorescence decay was measured in nitrogen atmosphere.

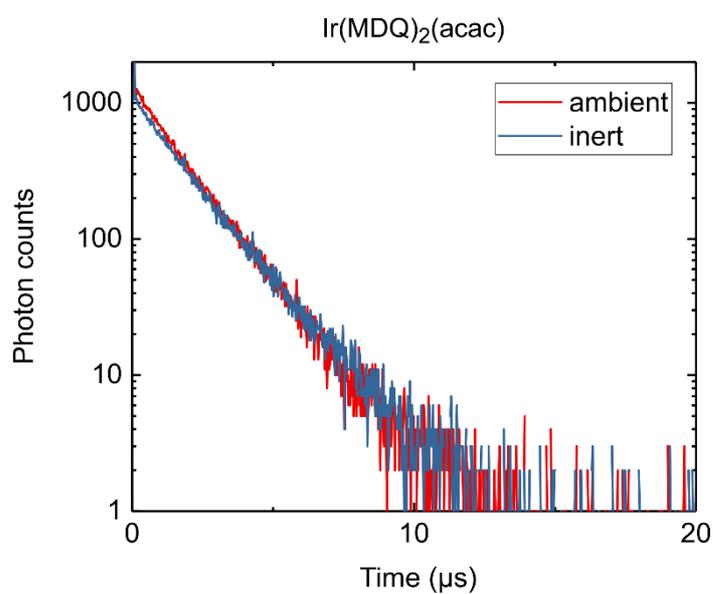

**Figure S6.** Phosphorescence decays of a PMMA:Ir(MDQ)$_2$(acac) (6 wt%) sample. The decays were measured in ambient and in inert nitrogen atmosphere.

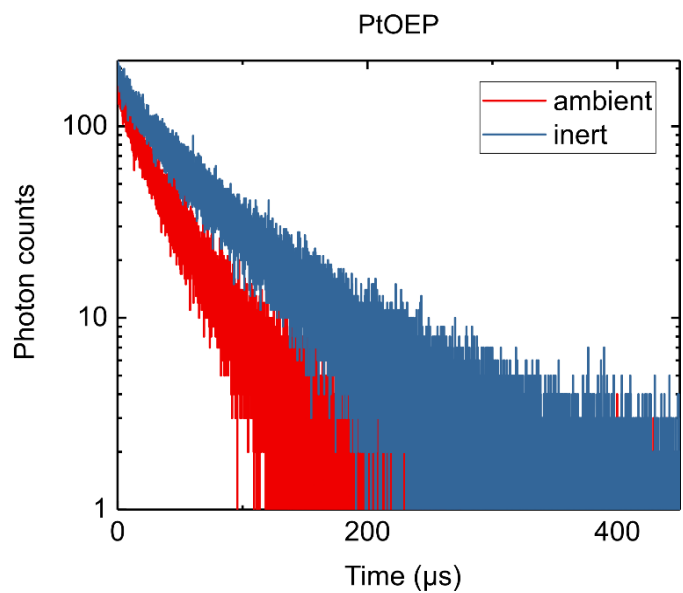

**Figure S7.** Phosphorescence decays of a PMMA:PtOEP (6 wt%) sample. The decays were measured in ambient and in inert nitrogen atmosphere.

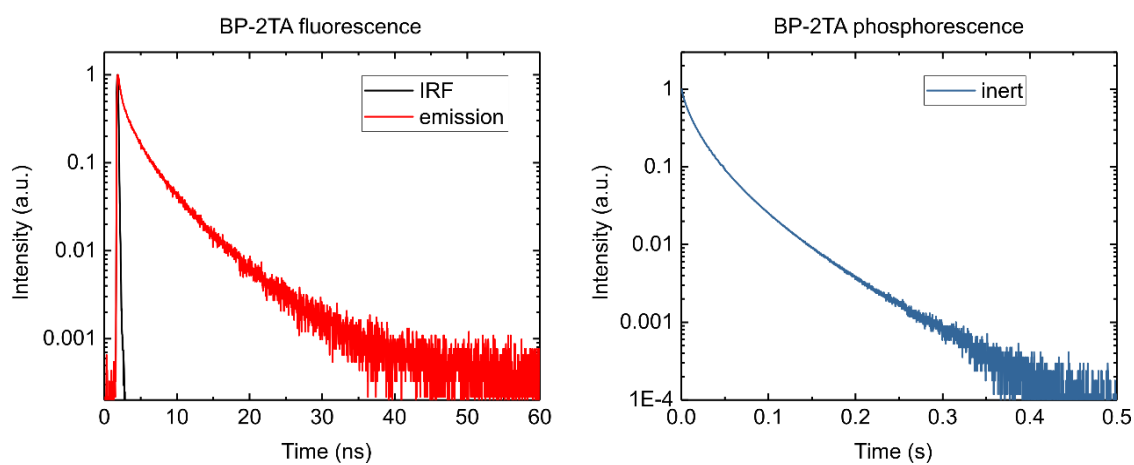

**Figure S8.** Fluorescence and phosphorescence decays of a PMMA:BP-2TA (5 wt%) sample. The fluorescence decay was measured in ambient atmosphere; the phosphorescence decay was measured in nitrogen atmosphere.